\documentstyle[aasms4]{article}

\begin{document}

\title{Self-similar temporal behavior of gamma-ray bursts}

\author{Andrei M. Beloborodov\altaffilmark{1,}\altaffilmark{2},
Boris E. Stern\altaffilmark{3}, Roland Svensson\altaffilmark{1}} 
\altaffiltext{1}{Stockholm Observatory, S-133 36, Saltsj{\"o}baden, Sweden} 
\altaffiltext{2}{Astro-Space Center of Lebedev Physical Institute, 
Profsoyuznaya 84/32, Moscow 117810, Russia} 
\altaffiltext{3}{Institute for Nuclear Research of Russian Academy of Science,
Profsojuznaja 7a, 117312 Moscow, Russia}

\begin{abstract}

We apply Fourier analysis to $214$ light curves of long gamma-ray bursts 
and study the statistical properties of their power density spectra (PDS). 
The averaged PDS is found to follow a power-law of index $-5/3$ over almost 
2 decades of frequency with a break at $\sim 2$ Hz. Individual PDS 
are exponentially distributed around the power-law. It provides evidence that 
the diversity of the bursts is due to random realizations 
of the same process which is self-similar over the full range of time-scales. 
The slope $-5/3$ of the average spectrum may indicate that gamma-ray bursts
are related to a 
phenomenon well studied in hydrodynamics -- fully developed turbulence.

\end{abstract}


\section{Introduction}

Discovered three decades ago, the phenomenon of gamma-ray bursts (GRBs)
remains
one of the mysteries of the Universe (Fishman \& Meegan 1995; Hartmann 1996). 
Being isotropically distributed on the sky (Meegan et al. 1996), the bursts 
are likely to occur at cosmological distances (Paczy\'nski 1992; Piran 1992; 
Kulkarni et al. 1998). Their light curves have many random peaks and in spite 
of extensive statistical studies (e.g., Nemiroff et al. 1994; Norris et al. 
1996; Stern 1996), the temporal behavior of GRBs remains a puzzle.

In our analysis we used light curves of GRBs observed by the Burst and 
Transient Source Experiment (BATSE) in the energy band, $50<h\nu<300$ keV, with 
64 ms resolution. 
The background was subtracted using linear fits to the 1024 ms data.
To avoid large Poisson fluctuations in the light curves,
we excluded dim bursts with peak count rates $<$ 250 counts per 64 ms bin.
(The background subtraction and the peak search are described in Stern, 
Poutanen, \& Svensson [1997, 1999]).
Long bursts are of particular interest as their internal temporal structure can
be studied by spectral analysis over a larger range of time-scales. We chose 
bursts with durations $T_{90}>20$ s where $T_{90}$ is the time it takes to 
accumulate from 5 \% to 95 \% of the total fluence of a burst. The resulting 
sample contains 214 bursts. We calculated the Fourier transform of each light 
curve, and found the corresponding power density spectrum (PDS) which is given 
by the squared Fourier amplitude summed for the two frequencies, $f$ and $-f$. 

Bursts have very diverse PDS, so that it is hard to see any systematic shapes 
in individual PDS (Giblin, Kouveliotou, \& van Paradijs 1998).
We therefore proceed studying the statistical properties of PDS 
for our sample of 214 bursts. In particular, we calculate the 
average PDS. This is meaningful if the bursts are produced by a common physical
mechanism, so that their light curves can be considered as time fragments of 
the same stochastic process. Such a process was suggested to be responsible for 
the stretched exponential shape of the average time profile of GRBs 
(Stern 1996; Stern \& Svensson 1996). 
We search for a signature for the underlying process in the PDS statistics.

\section{The average PDS}

The averaging of the PDS requires a specification of a weight for each burst.
We normalize GRB light curves setting their peak count rates to unity.
% (this is often considered as placing all bursts at a standard distance
% from the observer -- the ``standard candle'' normalization). 
This corresponds to the PDS normalization by the squared peak of the light 
curve. We will discuss 
below how a different normalization would affect the statistics of PDS. 

In Figure 1, the average power density spectrum, $\bar{P}_f$, for the sample of
214 peak-normalized bursts is presented. Poisson fluctuations of the time bin 
counts start to become important at high frequencies, $> 1$ Hz. 
The individual ``Poisson level'' of a burst equals its total fluence 
including the background. The spectrum above this level displays the 
intrinsic variability of the signal. The horizontal solid line in Figure 1 
shows the averaged normalized Poisson level. 

The striking feature of the average PDS is the power-law behavior over almost 
2 orders of magnitude in frequency. Its slope, $\alpha$, is approximately equal
to $\alpha=-5/3$  shown by the dashed line. The deviation from the power-law at
the low-frequency end is due to the finite duration of bursts (the average 
$T_{90}$ is about 80 s for our sample). To see the behavior of $\bar{P}_f$ at 
high frequencies, we smooth it on the scale $\Delta \log f = 0.03$ and multiply
by $f^{5/3}$. The result is shown by the solid curve in the top panel in 
Figure 2. The dotted curve displays the same spectrum after subtraction of 
the Poisson level. The subtraction was performed individually in each burst 
before the averaging. This is equivalent to subtraction of the effective level 
shown in Figure 1 from the average PDS. 

One can see that $\bar{P}_f$ is very close to a power-law with index
$\alpha=-5/3\approx -1.67$ in the range $0.02 < f < 1 $ Hz (the best power-law 
fit in this range has $\alpha = -1.67\pm 0.02$). After subtraction of the 
Poisson level, a break appears between 1 and 2 Hz. The existence of this break 
is better seen for bright GRBs, where we do not have to account for the Poisson 
noise. We therefore also plot the average PDS for the 27 bursts with peak count
rates $> 2000$ counts/bin (the bottom panel in Figure 2). The  break is again 
seen  at $\sim 2$ Hz. The break is too sharp to be explained as an artefact of 
64 ms time binning, which suppresses the PDS by a factor of
$[\sin(\pi f\Delta)/(\pi f \Delta)]^2$ where $\Delta=64$ ms is the time 
bin (cf. van der Klis 1989).

Is the power-law behavior of the average PDS related to the type of 
normalization we chose? To test this we tried other normalizations, e.g., by
dividing the light curve by the burst fluence. We found the same qualitative 
behavior of the average PDS: a power-law of index 
$\alpha\approx -5/3$ with super-imposed fluctuations. For
peak-normalization, the fluctuations in $\bar{P}_f$ are minimal 
and the best accuracy of the slope $-5/3$ is achieved. 
In this case, the amplitude of statistical fluctuations is consistent 
with the relation $\Delta \bar{P}_f/\bar{P}_f \sim N^{-1/2}$, where $N$ 
is the number of bursts in the sample.

\section{PDS distribution}

Power spectra of each individual burst are very diverse and they show strong 
deviations from the average power-law. The deviations, however, follow a simple 
statistical behavior. We constructed the corresponding histogram at each 
frequency and found that individual $P_f$ are distributed around $\bar{P}_f$ 
according to 
the law $dN/dP_f=N \exp(-P_f/\bar{P}_f)$. The histogram is not accurate as the 
number of bursts (N=214) is modest, but the statistics increases when we sum up 
the histograms at adjacent frequencies. After this summation, the $P_f$ 
distribution remains narrow being described by the exponential law as seen in 
Figure 3. Note that we get this distribution only when taking the bursts 
peak-normalized. For comparison, we also plot an analogous histogram for bursts 
normalized by fluence. 

The PDS of each individual burst can be decomposed into the power-law with 
$\alpha=-5/3$ and super-imposed exponentially distributed fluctuations. The 
exponential distribution,  also denoted the 2-dimensional $\chi^2$ 
distribution, just indicates that the two ($\sin$ and $\cos$) components of 
the Fourier transform are normally distributed around the average value 
(van der Klis 1989). Similar fluctuations are present in the PDS of the 
well-known standard noises such as Poisson noise (PDS slope $\alpha=0$), 
flicker noise ($\alpha=-1$), and Brownian motion ($\alpha=-2$). Note the 
difference of the GRB variability from the standard noises. In the case of a 
noise, the exponentially distributed fluctuations are suppressed when the PDS 
is smoothed by averaging $P_f$ over adjacent frequencies. By contrast, the 
smoothed power spectra of GRBs continue to show the exponential fluctuations 
around $\bar{P}_f$ independently of the smoothing scale. This behaviour makes 
it difficult to recognize the power-law in an individual burst. Nevertheless, 
the standard $P_f$ distribution around the mean slope $-5/3$ supports the view 
that different GRBs are random statistical realizations of the same stochastic
process described by the power-law.

\section{Discussion}

We conclude that the whole diversity of GRBs can be unified: each burst can be 
considered as a time fragment of the same process which is self-similar on 
time-scales from $\sim 0.5$ s up to the total duration of a burst. It follows 
that GRBs have a common origin being generated by some standard emission 
mechanism. Note, however, that only long bursts have been analyzed here, and 
short bursts may belong to a distinct class, as indicated by their duration 
distribution (Kouveliotou et al. 1993). The typical time-scale $\sim 0.5$ s 
observed in the PDS of long GRBs may be connected with the separate class of 
short, $\sim 1$ s, bursts.

Within the accuracy of measurement ($\sim 1$ \%), the slope of the average 
spectrum equals $-5/3$. Kolmogorov spectrum of velocity fluctuations in a 
turbulent medium has the same slope. Is it just a coincidence or is it an 
indication that bursts are somehow related to turbulence? In the cosmological 
scenario, the GRB emission is generated in a relativistic outflow from a 
central source (Rees \& M\'esz\'aros 1994) which is likely to result from 
the coalescence of two neutron stars or a neutron star and a black hole 
(Paczy\'nski 1986; Piran, Shemi, \& Narayan 1993). Turbulence develops easily 
in a super-sonic outflow and is known to occur, e.g., in the solar wind 
(Coleman 1968). Fluctuations of the velocity and magnetic field in the solar 
wind have PDS with slope $-5/3$, which indicates the presence of developed 
turbulence.

If GRBs are due to emission of bright blobs in a relativistic outflow, and the 
observed radiation is highly sensitive to the direction of the blob velocity, 
then a fluctuation of the velocity might "switch off" or "switch on" the blob 
in the observed light curve. This is only one of the possible speculations.  
A relation between GRB light curves and turbulence remains a conjecture until 
it is supported by a specific emission model. Additional information, which can 
be used in developing a model, is the 2 Hz break in the average PDS.
The break implies that pulses shorter than $\sim$0.5 s are suppressed in the 
light curve for some physical reason. The deficit of short pulses has also been
demonstrated directly by analyzing the pulse structure of the light curves 
(Norris et al. 1996). Note that the sharpness of the break supports the 
assumption of similar conditions in different GRBs. For example, if the Lorentz
factor of the emitting gas varied strongly from burst to burst we would observe
a smooth turnover instead of the sharp break.

\bigskip

We thank the referee for comments and A.F. Illarionov, L. Borgonovo, 
G. Bj\"ornsson, A.G. Doroshkevich, J. Poutanen and F. Ryde for discussions. 
This work was supported by the Swedish Natural Science Research Council,
the Swedish Royal Academy of Science, the Wennergren Foundation for 
Scientific Research, a NORDITA Nordic Project grant, and RFFI grant 97-02-16975.

\newpage
\section*{References}
\parindent=0pt

Coleman, P.\,J. 1968, ApJ, 153, 371

Fishman, G.\,J., \& Meegan, C.\,A. 1995, ARA\&A, 33, 415 
 
Giblin, T.\,W., Kouveliotou, C., \& van Paradijs, J. 1998,
in AIP Conf. Proc. 428, Gamma-Ray Bursts, \\
\hspace*{1cm} ed. C.\,A. Meegan, R.\,D. Preece, \& T.\,M. Koshut 
(New York: AIP), 241

Hartmann, D.\,H. 1996, A\&AS, 120, 4, 31

Kouveliotou, C., Meegan, C.\,A., Fishman, G.\,J., Bhat, N.\,P., Briggs, M.\,S., 
Koshut, T.\,M., Paciesas, W.\,S., \\
\hspace*{1cm} \& Pendleton, G.\,N.  1993, ApJ, 413, L101 

Kulkarni, S.\,R., et al. 1998, Nature, 393, 35

Meegan, C.\,A., et al. 1996, ApJS, 106, 65

Nemiroff, R.\,J., Norris, J.\,P., Kouveliotou, C., Fishman, G.\,J.,
Meegan, C.\,A., \& Paciesas, W.\,S.  1994, \\
\hspace*{1cm} ApJ, 423, 432

Norris, J.\,P., Nemiroff, R.\,J., Bonnell, J.\,T., Scargle, J.\,D.,
Kouveliotou, C., Paciesas, W.\,S., Meegan, C.\,A., \\
\hspace*{1cm} \& Fishman, G.\,J.  1996, ApJ, 459, 393

Paczy\'nski, B. 1986, ApJ, 308, L43

Paczy{\'n}ski, B. 1992, Nature, 355, 521

Piran, T. 1992, ApJ, 389, L45

Piran, T., Shemi, A., \& Narayan, R. 1993, MNRAS, 236, 861

Rees, M.\,J., \& M{\'e}sz{\'a}ros, P. 1994, ApJ, 430, L93

Stern, B.\,E. 1996, ApJ, 464, L11

Stern, B.\,E., \& Svensson, R. 1996, ApJ, 469, L109 

Stern, B.\,E., Poutanen, J., \& Svensson, R. 1997, ApJ, 489, L41

Stern, B.\,E., Poutanen, J., \& Svensson, R. 1999, ApJ, 510, in press

\noindent
van der Klis, M. 1989, in Timing Neutron Stars, ed. H. \"Ogelman \&
E.\,P.\,J. van den Heuvel, \\
\hspace*{1cm} NATO ASI C262, Kluwer, 27

\newpage       
\pagestyle{empty}
\parskip .6ex
\parindent 0mm
\section*{Figure captions}

{\bf Figure 1.} The averaged PDS for 214 peak-normalized bursts with duration 
$T_{90}>20$ s. The solid horizontal line shows the averaged Poisson level.

{\bf Figure 2.} {\it Top panel:} Same as Fig. 1, but now the average PDS is 
smoothed on a scale $\Delta\log f=0.03$ and multiplied by $f^{5/3}$ (solid 
curve). The dotted curve shows the spectrum after subtraction of the Poisson 
level. The error bars show the typical uncertainties in $\bar{P}_f$.
{\it Bottom panel:} The average PDS for the 27 brightest bursts in the sample.

{\bf Figure 3.} The solid histogram  displays the $P_f$ distribution of 214
peak-normalized GRBs. We divided the interval $-2.4<\log(P_f/\bar{P}_f)<1.2$ 
into 30 equal bins $\Delta\log (P_f/\bar{P}_f)$, and found how many bursts have
$P_f$ within a given bin. The histogram was constructed at each frequency and 
then summed up over all frequencies in the range $0.03 < f < 1$ Hz. The solid 
curve shows the exponential distribution, 
$dN/d\log P_f\propto P_f\exp(-P_f/\bar{P_f})$. The dotted histogram displays 
the $P_f$ distribution found when normalizing the light curves by fluence.
The integrals of all distributions are normalized to unity.

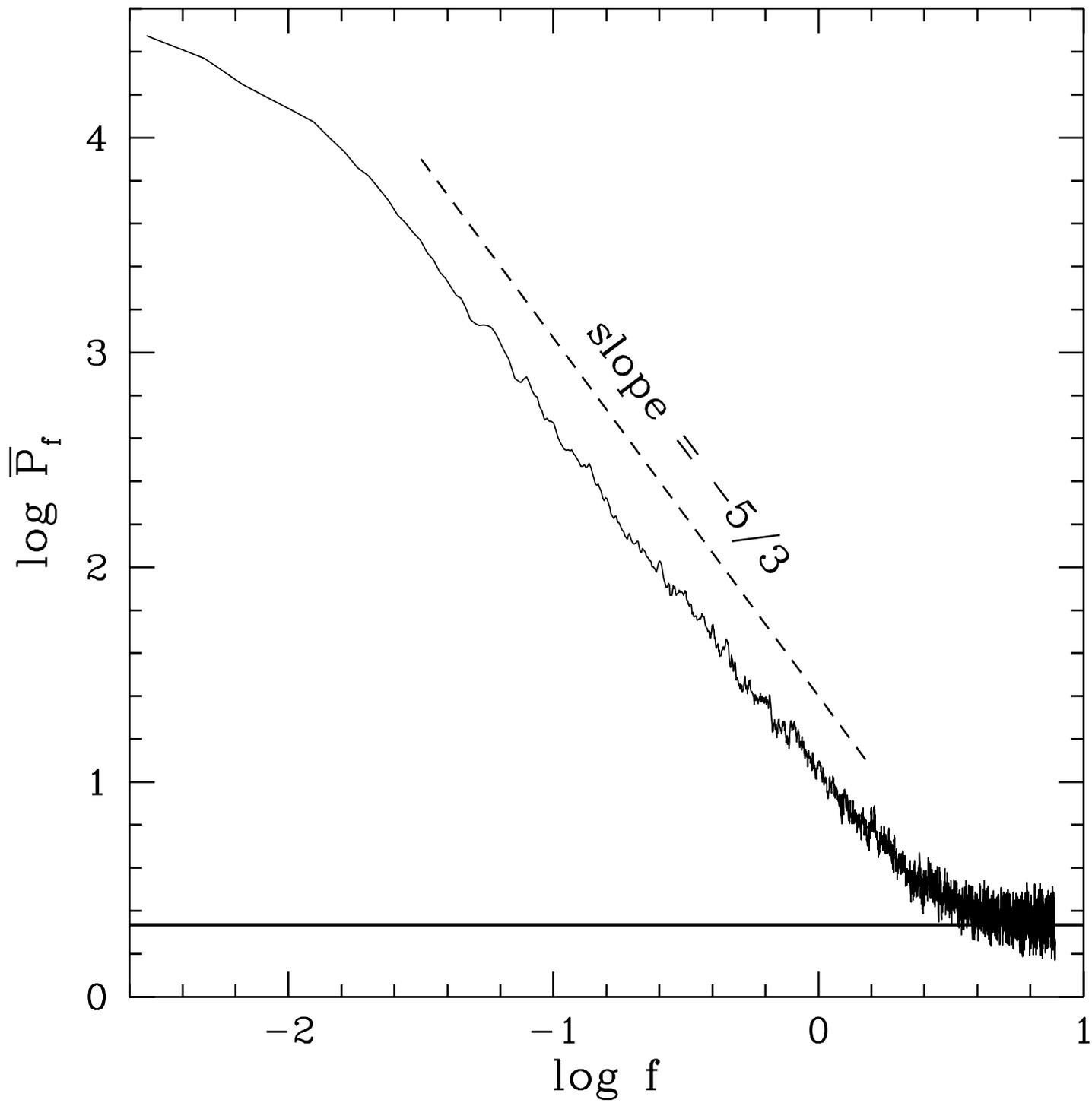
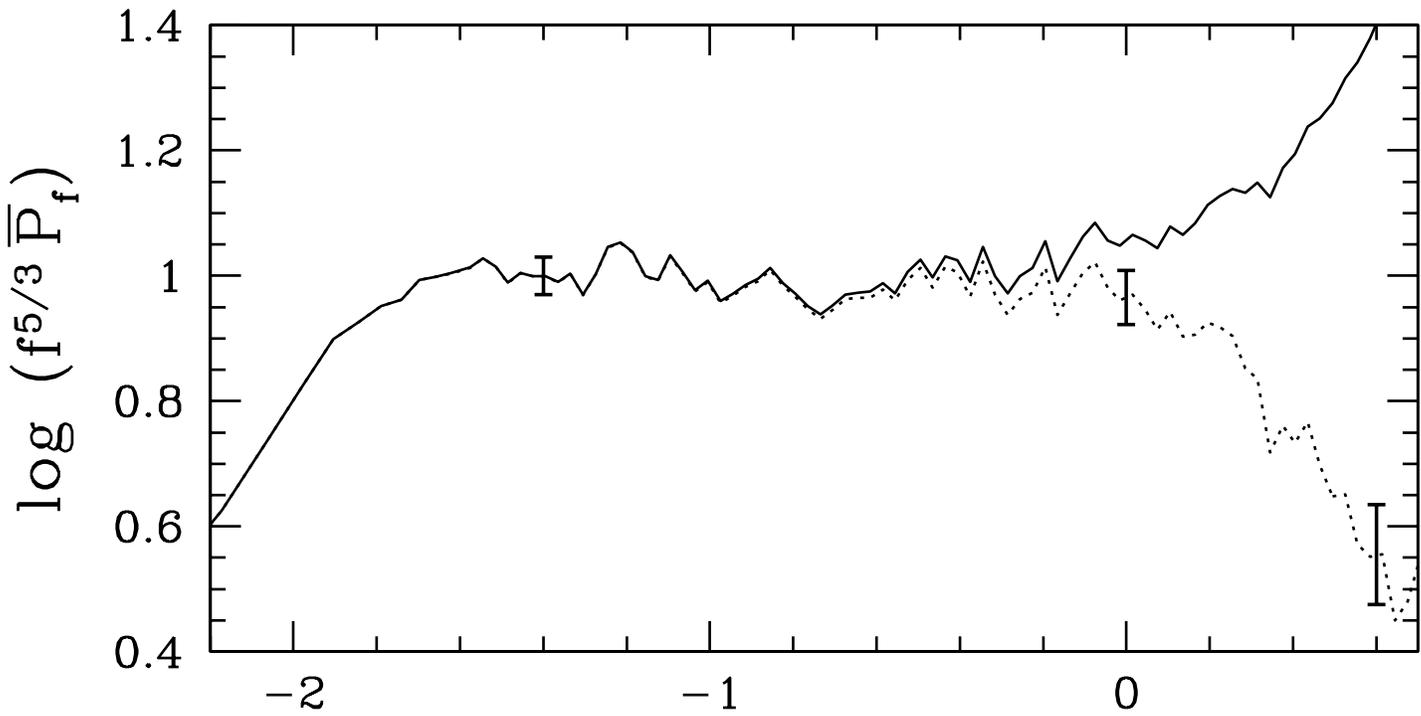
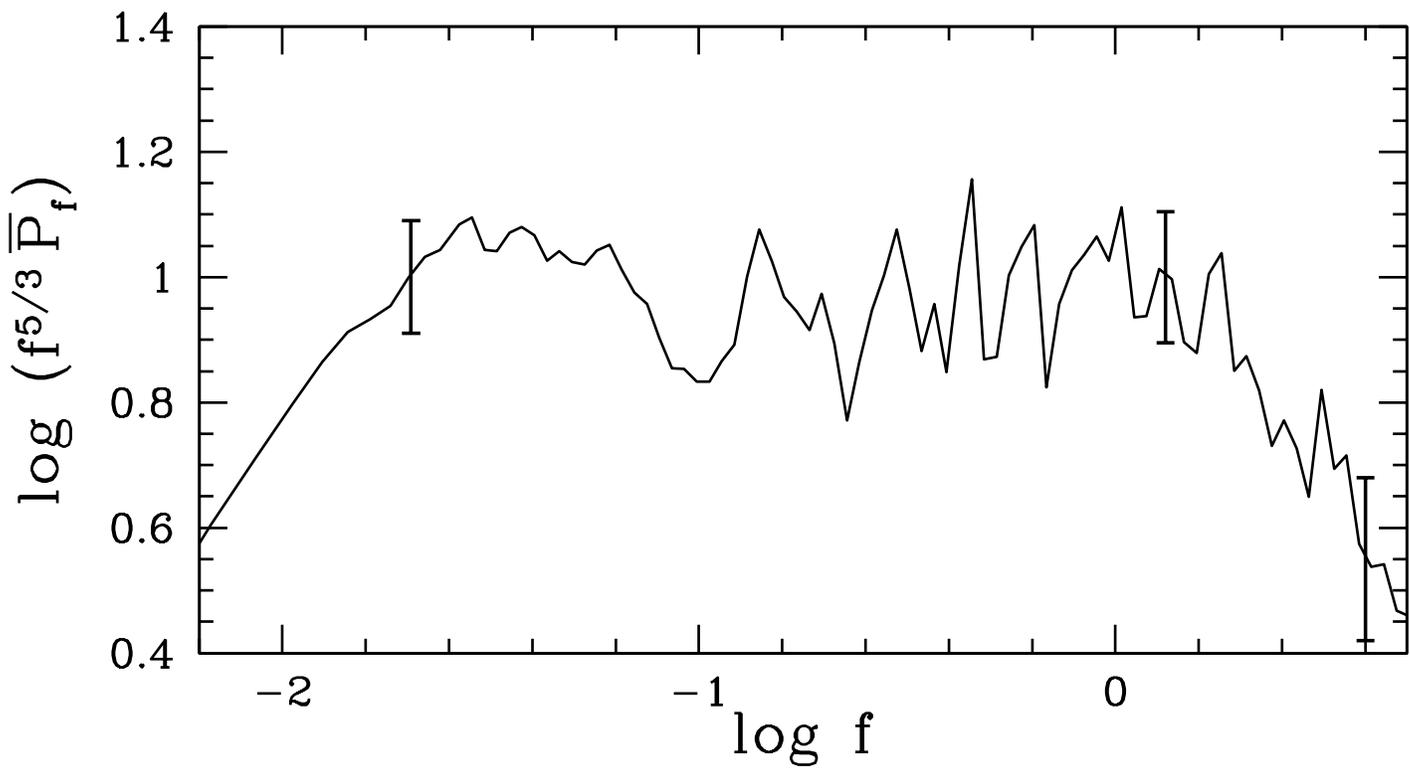
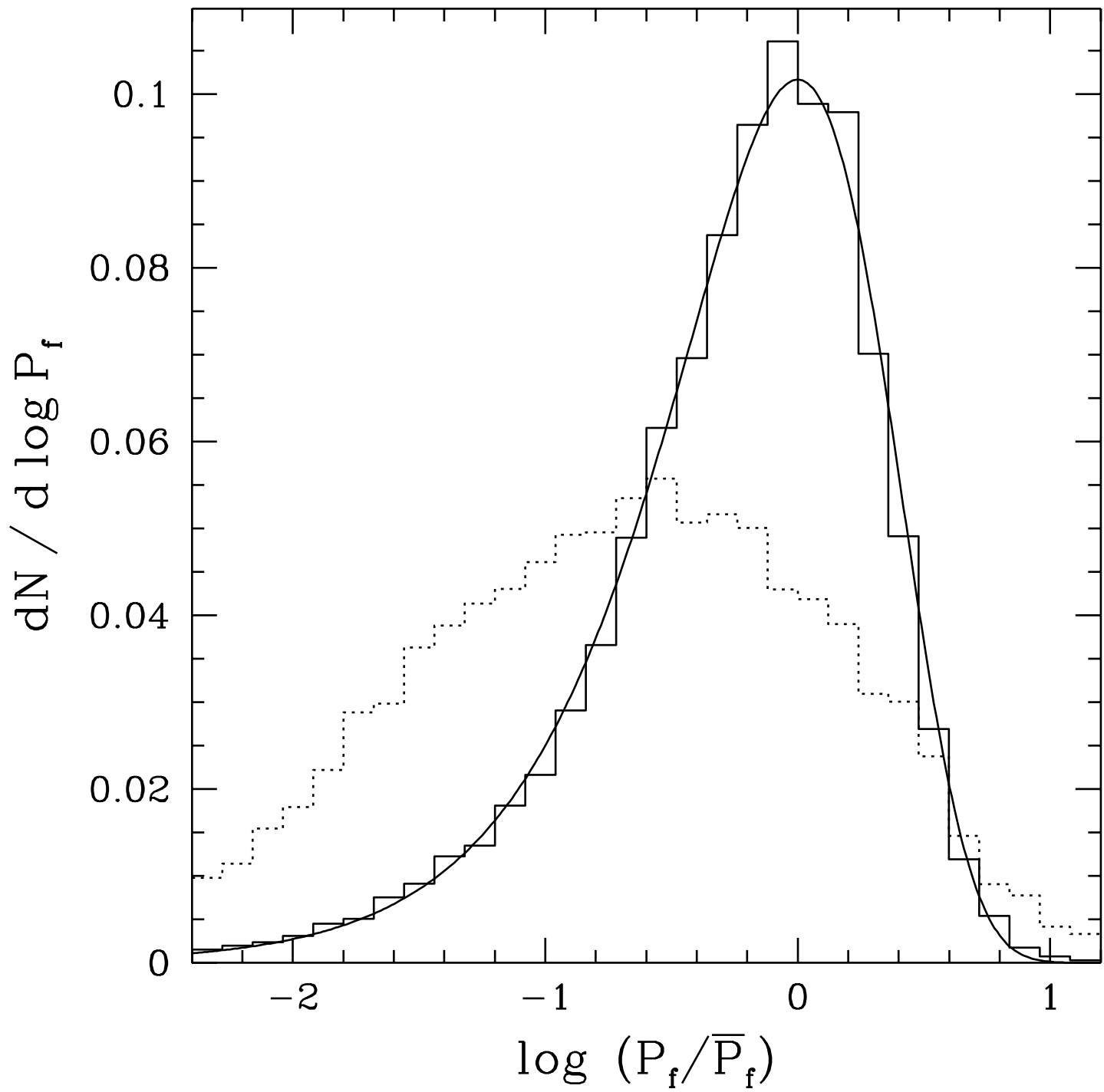
\end{document}